\documentclass[aps,nofootinbib,showpacs]{revtex4}
\begin{document}

\rightline{May 2005}

\title{
Avoiding the gauge heirarchy problem with see-sawed neutrino 
masses}

\author{R. Foot}\email{foot@physics.unimelb.edu.au}
\affiliation{School of Physics,
Research Centre for High Energy Physics,
The University of Melbourne, Victoria 3010, Australia}

\begin{abstract}

We show that the see-saw neutrino
mass mechanism can coexist naturally with an extended
gauge symmetry (i.e. without any gauge heirarchy problem) 
provided that the gauge symmetry contains 
gauged lepton number differences. The simplest such
`natural' see-saw models are constructed and their implications
for neutrino anomalies discussed.
\end{abstract}
\maketitle

%%%%%%%%%%%%%%%%%%%%%%%%%%%%%%%%%%%%%%%%%%%%%%%%%%%%%%%%%%%%%%%%%%%%%%%%%%%%%%%5

There is compelling evidence for non-zero neutrino masses
arising from the solar and atmospheric neutrino anomalies 
and various terrestrial experiments\cite{review2}.
A simple explanation of the required 
small neutrino masses, $m_{\nu} \stackrel{<}{\sim} eV$,
is provided through the see-saw mechanism\cite{seesaw}, which relies on a much
larger Majorana right-handed neutrino
mass scale $M \stackrel{>}{\sim} 10^6$ GeV.
Such a large scale would be problematic if it were generated by
the vacuum expectation value of a Higgs boson, since this would lead to
fine tuning in the Higgs potential, both 
at the tree-level and radiatively\cite{gil}.
However, this may not necessarily occur
since the right-handed neutrinos are electroweak gauge
singlets, which means that their mass scale might arise from
bare mass terms:
\begin{eqnarray}
{\cal L} = M_R^{ij}\bar \nu_{iR} (\nu_{jR})^c
\end{eqnarray}
If this were the case then 
there is no fine tuning problem in the theory at the
classical or tree-level. This would greatly alleviate
the gauge heirarchy problem.

Radiative corrections to the Higgs mass
still arise which are of order\cite{vis}:
\begin{eqnarray}
\delta \mu^2 \approx {\lambda^2 \over (2\pi)^2} M_R^2
log (\Lambda/M_R)
\end{eqnarray}
where $\lambda$ is the Higgs charged lepton Yukawa coupling
(and $\Lambda$ a momentum cutoff).
If we assume that
$\delta \mu^2 \stackrel{<}{\sim} TeV$ then
$M_R \stackrel{<}{\sim} 10^{7}-10^8$ GeV for neutrino
masses of order $10^{-1}$ eV\cite{vis}.
This suggests only a relatively low
scale for $M_R$ is `natural'.
Nevertheless, such
a relatively low scale for $M_R$ is still quite interesting
making such `natural' see-saw models worthy of study.

Our notation for the standard model quarks and leptons
under the standard gauge symmetry,
$SU(3)_c \otimes SU(2)_L \otimes U(1)_Y$, is
\begin{eqnarray} 
& f_{iL} & \sim (1, 2, -1), \ e_{iR} \sim (1, 1, -2)
\nonumber \\
& Q_{iL} & \sim (3, 2, 1/3), \ u_{iR} \sim (3, 1, 4/3),\ d_{iR} \sim (3, 1,
-2/3)
\end{eqnarray}
where $i=1,2,3$ is the generation index.

Our assumptions are
\begin{itemize}

\item Three right-handed neutrinos exist, which transform as
$\nu_{iR} \sim (1, 1, 0)$ under the standard model gauge symmetry.

\item The right-handed neutrinos transform non-trivially under
some extended gauge symmetry. 

\item We assume see-saw neutrino mass mechanism, with
the right-handed neutrino mass scale ( $\sim 10^7$ GeV) set 
by bare masses.
As explained above, this
is reasonable in order to avoid the gauge heirarchy problem. 
\end{itemize}
\noindent
Our task now is to derive the implications of
the above assumptions.

All gauge symmetries contain a $U(1)_{L'}$
subgroup and the assumption that the right-handed
neutrinos transform non-trivially means that at least one of the $\nu_{iR}$ 
has a non-zero $L'$ charge.
Define $L_e, L_{\mu}, L_{\tau}$ such that
$L_e f_{1L} = f_{1L},\ L_e e_{1R} = e_{1R}$ and $L_e F = 0$ for $F \neq f_{1L},
e_{1R}$.
$L_\mu, L_\tau$ are similarly defined in the obvious way.
Then the most general $U(1)$ that is a (classical) symmetry of the
standard model Lagrangian terms is generated by:
\begin{eqnarray}
L' = a L_e + b L_\mu + c L_\tau + \alpha L_{\nu_{1R}} + 
\beta L_{\nu_{2R}} + \gamma L_{\nu_{3R}} + \epsilon B + \omega Y
\end{eqnarray}
where $a, b, c, \alpha, \beta, \gamma, \epsilon, \omega$ are arbitrary
parameters and
$L_{\nu_{1R}}$ is the right-handed $\nu_{1R}$ number
(i.e. $L_{\nu_{1R}} \nu_{1R} = \nu_{1R}$ and $L_{\nu_{1R}} F = 0$
for $F \neq \nu_{1R}$). $L_{\nu_{2R}}$ and $L_{\nu_{3R}}$ are similarly 
defined in the obvious way. $B$ is the baryon number (defined
in the usual way with all quarks having charge 1/3).

In order to have light neutrinos without also having a gauge heirarchy problem, we require
all three bare Majorana right-handed neutrino 
mass eigenvalues to be non-zero. 
Define the right-handed neutrino bare mass matrix as follows:
\begin{eqnarray}
( \bar \nu_{1R}, \bar \nu_{2R}, \bar \nu_{3R}) \left(
\begin{array}{ccc} 
A & X & Y \\
X & B & Z \\
Y & Z & C 
\end{array}\right)
\left( \begin{array}{c}
(\nu_{1R})^c \\
(\nu_{2R})^c \\
(\nu_{3R})^c
\end{array}\right)
\end{eqnarray}
The bare masses must be $U(1)_{L'}$ invaraint.
If the $L'$ charge of at least
one of the $\nu_{iR}$ is non-zero, then without loss
of generality, we can assume that $L' \nu_{1R} \neq 0$,
which means that $A = 0$.
For $\nu_{1R}$ to have a bare mass, we require
either $X$ or $Y$ to be non-zero. This means that
the $L'$ charge of $\nu_{2R}$ and/or $\nu_{3R}$ 
must also be non-zero and have opposite $L'$ charge to $\nu_{1R}$.
For definiteness we assume $\nu_{2R}$ to have opposite
$L'$ charge to $\nu_{1R}$ which means that $A$ and $B$
are both zero and $X$ can be non-zero. If the $L'$ charge of $\nu_{3R}$
is non-zero, then $C=0$ and either $Y$ or $Z$ is also zero which
means that there is a zero eigenvalue.
Thus, we must have the $L'$ charge of $\nu_{3R}$ 
being zero. In other words, the requirement that
all three bare right-handed neutrino eigenvalues are
non vanishing together with the assumption that $L'$ charge
of at least one of the right-handed neutrinos is non-zero
leads uniquely to a right-handed Majorana mass matrix of the form:
\begin{eqnarray}
( \bar \nu_{1R}, \bar \nu_{2R}, \bar \nu_{3R}) \left(
\begin{array}{ccc} 
0 & X & 0 \\
X & 0 & 0 \\
0 & 0 & C 
\end{array}\right)
\left( \begin{array}{c}
(\nu_{1R})^c \\
(\nu_{2R})^c \\
(\nu_{3R})^c
\end{array}\right)
\end{eqnarray}
Not only is the form of the right-handed mass matrix unique 
but we must have
\begin{eqnarray}
\alpha = -\beta, \ \gamma = 0.
\label{c1}
\end{eqnarray}
(Obviously this is only unique up to 
trivial permutations of the $\nu_{iR}$ and $\alpha, \beta, \gamma$).

For quantum consistency, all gauge anomalies involving $U(1)_{L'}$
must vanish.
The anomaly cancellation\cite{abj,delb} conditions are:
\begin{eqnarray}
SU(2)^2_L U(1)_{L'} \ \Rightarrow  \ a + b + c + 3\epsilon = 0
\label{c2}
\end{eqnarray}
\begin{eqnarray}
U(1)_{L'}^3 \ \Rightarrow  \ \alpha^3 + \beta^3 + \gamma^3 - a^3
-b^3 - c^3 = 0 
\label{c3}
\end{eqnarray}
\begin{eqnarray}
{\rm mixed \ gauge-gravitational  \ anomaly}
\  \Rightarrow 
\ \alpha + \beta + \gamma - a - b - c = 0
\label{c4}
\end{eqnarray}
All other anomaly conditions do not give independent constraints.

Eq.(\ref{c1},\ref{c2},\ref{c3},\ref{c4}) imply that $\epsilon = 0$,
and either $a, b$ or $c$ is zero.
Thus, we find that the most general anomaly free form for $L'$, consistent
with a non-vanishing right-handed neutrino bare eigenmasses,
has the form:
\begin{eqnarray}
L' = c_1 \left( L_{\nu_{1R}} - L_{\nu_{2R}}\right) + c_2 \left( L_{e} -
L_{\mu} \right) + c_3 Y
\end{eqnarray}
where $c_1, c_2$ and $c_3$ are arbitary numbers. Obviously the form
is unique only up to permutations
of $\nu_{iR}$ and $L_{e,\mu,\tau}$.

So far we have not discussed how the extended gauge symmetry
is broken. We first examine the simplest case of just one
exotic Higgs multiplet, $h$. The most natural scale for the symmetry
breaking scale is $\langle h \rangle \stackrel{<}{\sim} TeV$
to avoid the gauge heirarchy problem. However, $\langle h \rangle$
cannot be too low, otherwise phenomenological problems
will arise. In particular,
if $h$ also breaks electroweak symmetry, then this would make
the $Z'$ light ($M_{Z'} \stackrel{<}{\sim} M_Z$) which
would be difficult to reconcile with existing experiments.
It is natural, therefore, to assume that $h$ is an electroweak
singlet.\footnote{Of course, this need only be the case if there
is one exotic scalar field. Later we will briefly consider
next to minimal models, with two exotic Higgs multiplets,
one of which can be an electroweak doublet. This doesn't
cause phenomenological problems if the vacuum expectation value (VEV) 
of the electroweak
singlet Higgs dominates over the electroweak doublet VEV's.}
This means that Dirac neutrino masses,
necessary for the see-saw mechanism to exist, must arise
through coupling with the standard model Higgs doublet, $\phi$:
\begin{eqnarray}
{\cal L} = \lambda_{ij} \bar f_{iL} \phi \nu_{jR}
\end{eqnarray}
This lagrangian is only $L'$ gauge invariant (with at least
two non-zero $\lambda's$) if
$|c_1| = |c_2|$ (and $c_1$ can be fixed to 1 without loss
of generality). This means that $L'$ can be taken as
\begin{eqnarray}
L' = L_{\nu_{1R}} - L_{\nu_{2R}} + L_e - L_\mu + \omega Y
\end{eqnarray}
or any of the other two physically distinct permutations\footnote{The phenomenological
implications of models with gauged lepton number differences has
been discussed previously in Ref.\cite{diff}.}.
Note that in the above basis, both the
Dirac neutrino mass matrix and charged lepton mass matrix are
both necessarily diagonal.

Since $L_e - L_\mu$ is a symmetry of the mass matrix
it follows that the effective Majorana left-handed neutrino
mass matrix has the form:
\begin{eqnarray}
( \bar \nu_{1L}, \bar \nu_{2L}, \bar \nu_{3L}) \left(
\begin{array}{ccc} 
0 & x & 0 \\
x & 0 & 0 \\
0 & 0 & y 
\end{array}\right)
\left( \begin{array}{c}
(\nu_{1L})^c \\
(\nu_{2L})^c \\
(\nu_{3L})^c
\end{array}\right)
\end{eqnarray}
This is the result in the absence of any coupling of the exotic Higgs
$h$ to the neutrino mass matrix. We see
that two flavours are maximally mixed but degenerate.
It is natural to assume that
the exotic Higgs $h$ couples to fermions 
(so that its gauge quantum numbers can 
be uniquely defined, c.f. ref.\cite{review}), which will induce 
corrections to the neutrino mass matrix.
There are just two possibilities (assuming $h$ is an electroweak
singlet), corresponding to $h$ having
gauge quantum numbers, $h \sim (1, 1, 0, +1)$ or $h \sim (1, 1, 0, +2)$ under
$SU(3)_c \otimes SU(2)_L \otimes U(1)_Y \otimes U(1)_{L'}$.
In the first case the right-handed Majorana mass matrix 
has additional terms coming from
\begin{eqnarray}
{\cal L} = \lambda \bar \nu_{1R} h (\nu_{3R})^c + 
\lambda' \bar \nu_{2R} h^* (\nu_{3R})^c  
+ H.c.
\label{14}
\end{eqnarray}
and so the right-handed Majorana mass matrix becomes:
\begin{eqnarray}
( \bar \nu_{1R}, \bar \nu_{2R}, \bar \nu_{3R}) \left(
\begin{array}{ccc} 
0 & X & \epsilon_1 \\
X & 0 & \epsilon_2 \\
\epsilon_1 & \epsilon_2 & C 
\end{array}\right)
\left( \begin{array}{c}
(\nu_{1R})^c \\
(\nu_{2R})^c \\
(\nu_{3R})^c
\end{array}\right)
\end{eqnarray}
where $\epsilon_1 = \lambda \langle h \rangle, \
\epsilon_2 = \lambda' \langle h \rangle$.

In the second case the right-handed Majorana mass matrix 
has additional terms coming from
\begin{eqnarray}
{\cal L} = \lambda \bar \nu_{1R} h (\nu_{1R})^c + 
\lambda' \bar \nu_{2R} h^* (\nu_{2R})^c  
+ H.c.
\label{16}
\end{eqnarray}
and the right-handed neutrino mass matrix has the form:
\begin{eqnarray}
( \bar \nu_{1R}, \bar \nu_{2R}, \bar \nu_{3R}) \left(
\begin{array}{ccc} 
\epsilon_1 & X & 0 \\
X & \epsilon_2 & 0 \\
0 & 0 & C 
\end{array}\right)
\left( \begin{array}{c}
(\nu_{1R})^c \\
(\nu_{2R})^c \\
(\nu_{3R})^c
\end{array}\right)
\end{eqnarray}

The second case features large angle
oscillations between two of the flavours with the
other flavour completely decoupled, 
while in the first case, we will have 
large angle oscillations between 2 flavours with the
other two angles being small.
Neither of these possibilities is compatible with the
currently popular 3-flavour solution to the neutrino anomalies.
If the popular neutrino solution pans-out then either
one of our three basic assumptions is incorrect, or
symmetry breaking is non-minimal (an example of non-minimal
symmetry breaking will be given later-on).

The first case might possibly be consistent with
experimental data if light sterile neutrinos exist.
In particular, the atmospheric neutrino anomaly could
be due to $\nu_\mu \to \nu_s$ oscillations\cite{atm1}\footnote{
The superKamiokande collaboration have argued\cite{sk} that
$\nu_\mu \to \nu_s$ oscillations are disfavoured relative
to the $\nu_\mu \to \nu_\tau$ possibility. However, as 
emphasised in Ref.\cite{footun}, the $\nu_\mu \to \nu_s$
hypothesis is still a possible solution
to the atmospheric neutrino anomaly. Long baseline experiments
will ultimately decide this issue.}
with solar oscillations being due to $\nu_e \to \nu_\tau$
oscillations. This scenario is potentially compatible
with the LSND experiment\cite{lsnd}, although the latter experiment
requires confirmation.
In order to incorporate the large angle $\nu_e \to \nu_\tau$
oscillations, this case 
would require the $L'$ symmetry
to be 
\begin{eqnarray}
L' = L_{\nu_{1R}} - L_{\nu_{3R}} + L_e - L_\tau + \omega Y
\label{last}
\end{eqnarray}
The neutral lepton mass matrix for $\nu_L, \nu_R$
has the form:
\begin{eqnarray}
{\cal L}_{eff} = {1 \over 2} ( \bar \nu_L \ \overline{(\nu_R)^c})
\left(
\begin{array}{cc}
0 & M_D \\
(M_D)^{T} & M_R 
\end{array}
\right)
\left(
\begin{array}{c}
(\nu_L)^c \\
\nu_R
\end{array}\right) \ + H.c.
\end{eqnarray}
In the see-saw limit where the eigenvalues of $M_R$ are much
larger than the eigenvalues of $M_D$, the right and
left neutrino states are effectively decoupled:
\begin{eqnarray}
{\cal L}^{see-saw} \simeq {1 \over 2} \bar \nu_L M_L (\nu_L)^c
+ {1 \over 2} \overline{(\nu_R)^c} M_R \nu_R
\end{eqnarray}
where
\begin{eqnarray}
M_L \simeq -M_D M_R^{-1} (M_D)^{\dagger}
\end{eqnarray}
In the case of Eq.(\ref{last}), $M_R$ is given by:
\begin{eqnarray}
M_R = 
\left(
\begin{array}{ccc} 
0 & \epsilon_2 & X \\
\epsilon_2 & C & \epsilon_1 \\
X & \epsilon_1 & 0 
\end{array}\right)
\end{eqnarray}
and $M_D$ is diagonal:
\begin{eqnarray}
M_D = \left( \begin{array}{ccc}
m_1 & 0 & 0 \\
0 & m_2 & 0 \\
0 & 0 & m_3
\end{array}\right)
\end{eqnarray}
Evaluating the effective light neutrino mass matrix, $M_L$, in
the limit where $\epsilon_{1,2} \ll X, C$, we
find it has the form:
\begin{eqnarray}
M_L = \left( \begin{array}{ccc}
0 & \omega_2 & y \\
\omega_2 & x & \omega_1 \\
y & \omega_1 & 0
\end{array}
\right)
\end{eqnarray}
where 
\begin{eqnarray}
x \simeq m_2^2/C,
\
y \simeq m_1 m_3/X,
\
\omega_1 \simeq -m_2 m_3 \epsilon_2/XC,
\
\omega_2 \simeq -m_1 m_2 \epsilon_1/XC.
\end{eqnarray}
Clearly we expect $\omega_i \ll x,y$ (note that the $\omega_i$ vanish in
the limit that $\epsilon_i \to 0$).
If the Dirac neutrino masses are heirarchial, it is natural
to expect $\omega_2 \ll \omega_1$, and we will examine
this limit for definiteness.
This matrix corresponds to the following oscillation
pattern:
\begin{eqnarray}
& {\rm Approximately} & {\rm \ maximal } 
\ \nu_e \leftrightarrow \nu_\tau \ {\rm oscillations} \ {\rm with}
\ \delta m^2 \simeq {2xy\omega_1^2 \over y^2 - x^2 }
\nonumber \\
&{\rm Small} \ {\rm angle} & \ \nu_{e}  \leftrightarrow \nu_\mu
\ {\rm 
oscillations \ with} \sin^2 2\theta_{e-\mu} = {4\omega_1^2 y^2 \over (x^2
- y^2)^2}, \ \delta m^2 \simeq x^2 - y^2 \nonumber \\
&{\rm  Small } \ {\rm  \ angle} & \ \nu_{\tau} \leftrightarrow \nu_\mu
\ {\rm 
oscillations \ with} \sin^2 2\theta_{\tau-\mu} = {4\omega_1^2 x^2 \over (x^2
- y^2)^2}, \ \delta m^2 \simeq x^2 - y^2 
\end{eqnarray}
If we apply the $\nu_e \leftrightarrow \nu_\tau$ oscillations
to the solar neutrino problem and the $\nu_e \leftrightarrow \nu_\mu$
oscillations to explain the LSND data, then we require:
\begin{eqnarray} 
y &\approx & \sqrt{|\delta m^2_{lsnd}|} \approx 0.5-1 eV
\nonumber \\
x &\approx & {2y \over \sin^2 2\theta_{lsnd}}{\delta m^2_{solar} \over 
\delta m^2_{lsnd}} \sim 10^{-1} \ eV
\nonumber \\
\omega_1 & \sim & 10^{-2} \ eV
\end{eqnarray}
There are two obvious problems with the above interpretation 
of the solar and LSND data.
First, the $\nu_e \leftrightarrow
\nu_\tau$ oscillations are predicted to be approximately maximal,
which gives a poor fit to the solar data: 
the hypothesis of maximal oscillations is
allowed by the SNO data at only about 1\% C.L. 
($\chi^2/d.o.f = 55.3/34$. \cite{solar}).
Second, the required value of $\omega_1$ is uncomfortably
large (or equivalently, this scheme would suggest a value of
$\sin^2 2\theta_{e-\mu} \ll 10^{-3}$, the value favoured
by the LSND experiment). 
Neither of these problems constitutes a rigourous experimental
exclusion of this scheme. Future data may well
exclude it but for now it is possible (but  
disfavoured by the data). 

To explain the atmospheric neutrino data, one can
assume the existence of at least one light sterile
neutrino maximally mixed with $\nu_\mu$ as in Ref.\cite{recent}.
This comes about naturally if 
a mirror sector exists\cite{flv,flv2}. Only
$\nu_{\mu L}, \nu_{2R}$ couple to their mirror
partners because only
these particles have zero $L'$ charge.
This leads to an effective mass matrix
for the 3 light ordinary and 3 light mirror
neutrinos of the form:
\begin{eqnarray}
M_L = 
\left( \begin{array}{cccccc}
0 & 0 & y & 0 & 0 & 0 \\
0 & x & \omega_1 & 0 & \delta & 0 \\
y & \omega_1 & 0 & 0 & 0 & 0\\
0 & 0 & 0 & 0 & 0 & y \\
0 & \delta & 0 & 0 & x & \omega_1 \\
0 & 0 & 0 & y & \omega_1 & 0 
\end{array}\right)
\end{eqnarray}
where the $\delta$ term is the effective $\bar \nu_{\mu L}
(\nu_{\mu L})^c$ mass mixing caused by the coupling of
the ordinary and mirror sectors.
The effect of the $\delta $ is to cause maximal
oscillations between the ordinary and mirror neutrinos
with the largest $\delta m^2$ occuring for 
$\nu_\mu \leftrightarrow \nu'_{\mu}$ oscillations.
In fact we find $\delta m^2_{atm} \simeq 4 x \delta$.
The maximal oscillations between $\nu_e \leftrightarrow
\nu'_e$ and $\nu_\tau \leftrightarrow \nu'_{\tau}$
are governed by, $\delta m^2_{11}, \delta m^2_{33}$
satisfying:
\begin{eqnarray}
\delta m^2_{11} &\approx & \delta m^2_{33} \approx
{2 \omega_1^2 \delta \over y} 
\end{eqnarray}
If the $\nu_e \leftrightarrow \nu_\mu$ oscillation
interpretation of the LSND experiment is correct, then
\begin{eqnarray}
\delta m^2_{11} \approx \sin^4 \theta_{lsnd}{\delta m^2_{lsnd}\over 16} {\delta m^2_{atm}
\over \delta m^2_{solar}}
\sim 10^{-5} \ eV^2
\end{eqnarray}
This would imply a much larger solar boron neutrino flux then
predicted by standard solar models (c.f. Ref.\cite{barger}). Alternatively, if the 
LSND experiment is not confirmed then we can have $\sin^2
2\theta_{e-\mu} \ll 10^{-3}$ and $\delta m^2_{11} \approx \delta m^2_{33}
\stackrel{<}{\sim} 10^{-11}\ eV^2$, implying that the $\nu_e
\leftrightarrow \nu'_e$ and $\nu_\tau \leftrightarrow \nu'_\tau$
oscillations would have no effect for the solar neutrino experiments.

So far, we have focussed on the minimal see-saw model
without gauge heirarchy problem. 
The minimal model has the theoretical advantage of
making definite predictions for the structure of the neutrino
mass matrix. The disadvantage is that the minimal model is 
experimentally disfavoured by a 
variety of experiments, although it is 
not yet definitely ruled out.  
If future experiments rule out the minimal model (by e.g.
confirming the $\nu_\mu \to \nu_\tau$ oscillation interpretation
of the atmospheric neutrino anomaly), then
it means that at least two exotic Higgs multiplets are required.
Unfortunately, this makes possible scenarios much more
(theoretically) arbitrary. For illustration, we give one 
relatively simple model, which we now outline.  

We assume that the gauge symmetry contains
$L' = L_e - L_\mu + L_{\nu_{1R}} - L_{\nu_{2R}}$, which is
broken by two exotic Higgs multiplets $h_1$ and $h_2$. The $h_1$ is an
electroweak singlet, with VEV $\sim$ TeV. The $h_1$ may couple to
the right-handed neutrino sector, either by Eq.(\ref{14}) or
Eq.(\ref{16}). The $h_2$ is an electroweak doublet,
$h_2 \sim (1, 2, +1, +1)$ under the gauge group,
$SU(3)_c \otimes SU(2)_L \otimes U(1)_Y \otimes U(1)_{L'}$.
This means that $h_2$ couples in the following way to leptons:
\begin{eqnarray}
{\cal L} = 
\lambda_1 \bar f_{1L} h_2 e_{3R} + \lambda_2 \bar f_{3L} h_2 e_{2R} 
+ \lambda_3 \bar f_{2L} h_2^c \nu_{3R} + \lambda_4 \bar f_{3L} h_2^c \nu_{1R} 
+ H.c.
\end{eqnarray}
In the model where $h_1$ couples via Eq.(\ref{14}), so that
$h_1 \sim (1, 1, 0, +1)$, the Higgs field $h_2$ can gain a naturally small
VEV, via terms such as $m h_1^c \phi^{\dagger} h_2$ in the Higgs
potential. Anyway, the point of $h_2$ is that the Dirac mass matrix
for the charged leptons and neutrinos is no longer necessarily diagonal.
The charged lepton mass matrix has the form:
\begin{eqnarray}
( \bar e'_{1L}, \bar e'_{2L}, \bar e'_{3L})
\left(
\begin{array}{ccc} 
m_1 & 0 & a \\
0 & m_2 & 0 \\
0 & b & m_3
\end{array}\right)\left(
\begin{array}{c}
e'_{1R} \\
e'_{2R} \\
e'_{3R}
\end{array}\right)
\end{eqnarray}
where $a = \lambda_1 \langle h_2 \rangle, \
b = \lambda_2 \langle h_2 \rangle$.
This mass matrix can accomodate maximal mixing in the left-handed 2-3 block
in the limit $a \to 0, \ |m_2 b| \gg |m_3^2 + b^2 - m_2^2 |$.
If we assume that
$\lambda_{3,4}$ are small so that the neutrino mass matrix 
is diagonalized by approxiately maximal mixing in the 1-2 sector, then the
effective light neutrino mixing matrix becomes approximately:
\begin{eqnarray}
K = 
\left( 
\begin{array}{ccc}
1 & 0 & 0 \\
0 & 1/\sqrt{2} & 1/\sqrt{2} \\
0 & -1/\sqrt{2} & 1/\sqrt{2}
\end{array}\right)
\left( 
\begin{array}{ccc}
1/\sqrt{2} & -1/\sqrt{2} & 0 \\
1/\sqrt{2} & 1/\sqrt{2} & 0 \\
0 & 0 & 1
\end{array}\right)
\end{eqnarray}
That is, bimaximal mixing results\cite{bimax}. Of course, course this is not
uniquely predicted and deviations can easily
occur (and are suggested by the data). 
In fact, any solar angle can be accomodated if $\lambda_{3,4}$
are non-zero.
Certainly, it would be interesting to do a detailed
systematic study of next to minimal models, something
we leave for the future\footnote{
One interesting feature of next to minimal models is that
we can embed the $U(1)_{L'}$ into a simple group such as $SU(2)$.
That is, we have gauge symmetry: $SU(3)_c \otimes SU(2)_L
\otimes U(1)_Y \otimes SU(2)$.
[This is not possible in the minimal model with only one exotic
Higgs multiplet because they have phenomenologically unacceptable
mass relations for the charged leptons].
Embedding the $U(1)_{L'}$ into simple groups such as $SU(2)$ may 
lead to more predictive and therefore more interesting models.
}.

To conclude this work, we have shown that the see-saw neutrino
mass mechanism can coexist naturally with an extended
gauge symmetry, without any gauge heirarchy problem, 
provided that the gauge symmetry contains 
gauged lepton number differences. The minimal model of this type
was examined, which was shown to give definite predictions
for the neutrino oscillation pattern which is theoretically very
interesting despite the relatively poor agreement of the 
derived model with the current experimental data. We have also
shown that the next to minimal models can accomodate
a wider spectrum of oscillation patterns, including approximately
bimaximal neutrino oscillations.

\vskip 0.5cm
\noindent 
{\bf Acknowledgements} 
The author would like to thank R. R. Volkas for his 
comments on a draft of this paper and for checking some
of the equations. Comments by K. McDonald are also greatfully
acknowledged. This work was supported by the
Australian Research Council.


\begin{thebibliography}{99}

\bibitem{review2}
For a recent review, see e.g.  K. Zuber, hep-ex/0502039.

\bibitem{seesaw}
P. Minkowski, Phys. Lett. B67, 421 (1977);
M. Gell-Mann, P. Ramond and R. Slansky,
in {\it Supergravity} (North Holland 1979) 315;
T. Yanagida in {\it Proceedings of the Workshop
on Unified Theories and Baryon Number in the
Universe} (KEK 1979).

\bibitem{gil}
E. Gildener, Phys. Rev. D14, 1667 (1976).

\bibitem{vis}
F. Vissani, Phys. Rev. D57, 7027 (1998).
%   hep-ph/9709409.


\bibitem{abj}
S. L. Adler, Phys. Rev. 177, 2426 (1969);
J. S. Bell and R. Jackiv, Nuovo Cimento
A60, 49 (1969).


\bibitem{delb}
R. Delbourgo and A. Salam, Phys. Lett.
B40, 381 (1972); T. Eguchi and P. Freund,
Phys. Rev. Lett. 37, 1251 (1976);
L. Alvarez-Gaume and E. Witten,
Nucl. Phys. B234, 269 (1983).


\bibitem{diff}
X-G. He, G. C. Joshi, H. Lew and R. R. Volkas,
Phys. Rev. D44, 2118, (1991).


\bibitem{review}
R. Foot, H. Lew and R. R. Volkas, J. Phys. G19, 361 (1993).

\bibitem{atm1}
R. Foot, R. R. Volkas and O. Yasuda, Phys. Rev. D58, 013006 (1998).

\bibitem{sk}
Super-Kamiokande Collaboration, S. Fukuda {\it et al}.,
Phys. Rev. Lett. 85, 3999 (2000).

\bibitem{footun}
R. Foot, Phys. Lett. B496, 169 (2000); Mod. Phys. Lett. A18, 2071
(2003).

\bibitem{lsnd}
LSND Collaboration, C. athanassopoulos {\it et al}.,
Phys. Rev. Lett. 81, 1774 (1998); Phys. Rev. C58, 2489
(1998); Phys. Rev. D64, 112007 (2001).

\bibitem{solar}
A. Bellerive (for SNO collaboration), hep-ex/0401018.

\bibitem{recent}
R. Foot and R. R. Volkas, Phys. Lett. B543, 38 (2002);
R. Foot, Mod. Phys. Lett. A18, 2079 (2003).

\bibitem{flv}
R. Foot, H. Lew and R. R. Volkas,
Phys. Lett. B272, 67 (1991). See also,
T. D. Lee and C. N. Yang, Phys. Rev. 104, 256 (1956);
I. Kobzarev, L. Okun and I. Pomeranchuk, Sov. J. Nucl. Phys. 3,
837 (1966); M. Pavsic, Int. J. Theor. Phys. 9, 229 (1974).

\bibitem{flv2}
R. Foot, H. Lew and R. R. Volkas, Mod. Phys. Lett. A7, 2567 (1992);
R. Foot, Mod. Phys. Lett. A9, 169 (1994); R. Foot and R. R. Volkas,
Phys. Rev. D52, 6595 (1995).


\bibitem{barger}
V. Barger {\it et al}., Phys. Lett. B537, 179 (2002).

\bibitem{bimax}
V. Barger {\it et al}., Phys. Lett. B437,
107 (1998); A. Baltz, A. S. Goldhaber and 
M. Goldhaber, Phys. Rev. Lett. 81, 5730 (1988);
F. Vissani, hep-ph/9708483;
D. V. Ahluwalia, Mod. Phys. Lett. A13, 2249 
(1988).



\end{thebibliography}
\end{document}